\documentstyle[12pt]{article}
\begin{document}
\title{{\it Sous les Chemises, la Sym\'etrie... }}
\author{Serge Galam \\ 
Laboratoire des Milieux D\'{e}sordonn\'{e}s 
et H\'{e}t\'{e}rog\`{e}nes\thanks{Laboratoire associ\'{e} au 
CNRS (UMR n$^{\circ}$ 7603)},\\
Tour 13 - Case 86, 4 place Jussieu, \\ 75252 Paris Cedex 05, France\\ 
(galam@ccr.jussieu.fr)
}
\date{Pour La Science (Paris) \\Hors s\'erie \\``Les Sym\'etries du Monde", 
\\ $\ $\\ 16-19, Juillet 1998}

\maketitle
\baselineskip 24pt

\newpage
%%%%%%%%%%%%%%%%%%%%%%%%

\begin{center}
 L'apparition spontan\'ee de l'ordre dans la mati\`ere r\'esulte de brisures
de sym\'etrie. Comment l'\'etat collectif d'un syst\`eme passe-t-il d'une sym\'etrie 
\`a une autre, comment brise-t-il sa sym\'etrie ?
C'est ce que nous allons voir, \`a travers une m\'etaphore de chemises
et de couleurs.

\end{center}

Le ph\'enom\`ene de brisure spontan\'ee de sym\'etrie  est un m\'ecanisme essentiel
de la physique des comportements collectifs dans la mati\`ere. 
Il est 
\`a l'origine de l'existence de structures ordonn\'ees, \`a partir desquelles 
apparaissent de nombreuses propri\'et\'es physiques, inexistantes
au niveau de l'atome ou de la mol\'ecule isol\'ee. Il marque une transition entre
un \'etat collectif de sym\'etrie \'elev\'ee (comme un liquide, sym\'etrique par toutes 
translations et rotations continues)
et un \'etat 
ordonn\'ee, de sym\'etrie moindre (comme un cristal, sym\'etrique par 
rapport \`a seulement certaines translations et rotations discontinues).
Citons par exemple, l'aimantation
d'un syst\`eme magn\'etique (la capacit\'e d'un aimant \`a attirer un clou), 
la supraconductivit\'e d'un alliage (la conduction d'\'electricit\'e sans 
aucune perte d'\'energie), 
et aussi peut-\^etre, la cr\'eation
de l'univers (s\'eparation de la mati\`ere de l'anti-mati\`ere, lors du Big Bang).

Plus commun\'ement, la brisure spontan\'ee de sym\'etrie est responsable des 
aspects tr\`es diff\'erents, sous lesquels on peut trouver une m\^eme 
et unique substance : ainsi, l'eau peut \^etre liquide, solide quand elle est
 glace, ou 
gazeuse quand elle est vapeur. Autre exemple : 
le fer peut \^etre en phase paramagn\'etique (les moments magn\'etiques 
atomiques sont d\'esordonn\'es), ou dans la phase ferromagn\'etique de l'aimant 
(tous les moments magn\'etiques atomiques pointent dans la m\^eme direction).  

Pour d\'ecrire
les brisures de sym\'etrie et leur dynamique,
 utilisons une m\'etaphore sociale : au lieu d'atomes portant des moments
magn\'etiques, imaginons des individus portant des chemises de couleurs diff\'erentes.

La chemise fait le moine.
 
Observons le comportement d'une population de N personnes, chacune isol\'ee 
dans une chambre, 
avec \`a sa disposition une chemise verte et une rouge.
Chaque personne choisit la chemise de son gožt, sans connaitre le choix des
autres. Vu de l'ext\'erieur, ce choix est al\'eatoire
comme l'est le r\'esultat pile ou face 
du lancer d'une pi\`ece de monnaie. 
Lorsque chacun a fait son choix, nous obtenons
une configuration particuli\`ere de la distribution des couleurs. Chacun 
pouvant faire deux choix, ind\'ependamment du choix des autres, 
nous avons $2^{N}$ configurations possibles. Ce nombre croit tr\`es vite 
avec le nombre de personnes N. D'une dizaine (16) pour N=4, il d\'epasse d\'ej\`a 
le million  (1 048 576) pour N=20 (songez qu'il y a environ $10^{22}$ molecules
dans un gramme d'eau).

Si N est assez grand, la plupart des configurations ont le m\^eme nombre 
de chemises rouges et vertes (bien que chaque configuration soit 
\'equiprobable). L'\'etat obtenu est sym\'etrique en moyenne,
par rapport \`a l'\'echange des couleurs verte et rouge.

Une fois les choix individuels r\'ealis\'es, nous supprimons tous les murs :
chaque personne voit la couleur de la chemise de ses
voisins, et seulement de ses voisins. Supposons maintenant que chaque individu 
ait une tendance extr\`eme \`a l'imitation, et repla\c cons les murs pour permettre
\`a chacun de refaire un choix de couleur : chaque individu souhaite porter
la couleur de chemise que la majorit\'e de ses voisins portait, et \`a partir 
de ce qu'il a vu, d\'ecide de
garder la m\^eme chemise, ou d'en changer. 
Nous retirons de nouveau les murs, et 
tout un chacun observe ses voisins. Puis nous remettons les murs, et une
autre r\'eactualisation de couleur se fait, et ainsi de suite un certain 
nombre de fois.

Nous imaginons assez bien que l'on va assister \`a des changements, apparemment
d\'esordonn\'es, et intempestifs de couleur de chemises. Mais au bout d'un certain 
temps, soudainement, comme par miracle, tout le monde va porter la m\^eme couleur. 
\` A ce stade, le ph\'enom\`ene est similaire \`a l'effet de mode bien connu 
en sociologie.
Mais ce processus a toutefois quelque chose de surprenant, car au d\'ebut, on va 
voir se former de ci de l\`a, des petits groupes homog\`enes d'une seule couleur,
certains en rouge, d'autres en vert.
Mais localement, \`a la fronti\`ere de ces groupes, il y aura 
comme des luttes d'expansion d'un
groupe au d\'etriment d'un autre, tant™t le rouge, tant™t le vert.
Cependant, petit \`a petit, in\'exorablement, 
un de ces petits groupes, apparemment comme les autres, va se mettre \`a croitre, 
au del\`a de la taille moyenne des autres groupes, et cela,
sans qu'on sache trop
pourquoi. Mais alors, il va
s'\'etendre, et tr\`es vite, de fa\c con irreversible 
\`a toute l'assembl\'ee. D'un coup tout le monde portera la m\^eme couleur,
et n'en changera plus.

Si l'exp\'erience est  r\'ep\'et\'ee, on retrouvera le m\^eme effet d'une couleur unique,
mais la couleur collectivement choisie sera tant™t le rouge, tant™t le vert.
On constate donc, \`a ce stade, que l'interaction visuelle entre voisins uniquement,
suffit \`a  produire 
un ordre de couleur, qui s'\'etend \`a l'ensemble du groupe,
donc bien au del\`a des proches
voisins que chacun voit. 
Mais simultan\'ement, le choix de la couleur semble arbitraire. Parfois
c'est le vert qui l'emporte,
et parfois c'est le rouge.

Ce ph\'enom\`ene d'uniformisation est ce qu'on appelle en physique 
une brisure spontan\'ee de sym\'etrie. 
Elle se produit d\`es que l'on introduit des interactions \`a courte port\'ee,
entre \'el\'ements de m\^eme nature, par exemple, 
dans un syst\`eme
de moments magn\'etiques (qu'on appelle des spins). \` A temp\'erature nulle, 
ceux-ci vont tous s'aligner parall\'element (c'est le ferromagn\'etisme)
sans exception, mais
dans une direction, elle, prise au hasard.
On a donc un effet de propagation spontan\'e d'un ordre \`a longue distance,
dans la direction des spins, alors que
les interactions entre ces m\^emes spins, ne sont qu'\`a courte distance.

C'est en fait, pour diminuer son \'energie interne,
que le syst\`eme dans son ensemble, va
s\'electionner un seul \'etat au hasard, au d\'etriment de tous les autres 
\'etats possibles. 
L'\'etat collectif est donc ``totalitaire", avec un seul et unique \'etat pour
tout le monde, sans aucune exception. Dans notre exemple de chemises,
lorque la sym\'etrie de couleur est bris\'ee, on n'y voit que du vert, ou
que du rouge.

En physique, cela n'est cependant vrai, que si la temp\'erature du syst\`eme 
reste nulle. 
D\`es qu'elle est diff\'erente de z\'ero, les choses se compliquent, avec
une r\'eintroduction partielle d'\'etats, auparavant exclus.
La temp\'erature va favoriser un ``non-conformisme individuel" avec
une augmentation des \'energies locales correspondantes. Un spin
va pouvoir ainsi s'orienter dans une direction diff\'erente de celle de 
la majorit\'e de ses
voisins. On dit en physique que le spin est ``excit\'e". 

La possibilit\'e, pour un spin d'\^etre excit\'e, est de nature probabiliste.
C'est \`a dire qu'\`a
chaque configuration locale dans l'orientation respective de ses voisins 
est associ\'ee 
une probabilit\'e d'occurence dont l'amplitude d\'epend de la valeur 
de la temp\'erature. Celle-ci
introduit donc
la
possibilit\'e statistique d'avoir de ci de l\`a, une configuration locale
de spins dont les orientations respectives  non seulement ne  minimisent pas
l'\'energie interne, mais au contraire peuvent m\^eme la maximiser.
Cela dans le cadre Boltzmanien d'une description statistique 
de la temp\'erature.

La notion de probabilit\'e introduit une dynamique dans la
distribution de ces configurations d'\'energie non-minimis\'ee, leur
dur\'ee de vie \'etant alors finie. Cela signifie
que ce ne sont pas toujours les m\^emes spins qui sont excit\'es au cours du temps. 
Le d\'esordre se d\'eplace, il est ``mobile".  
Il va donc y avoir une comp\'etition constante
entre l'ordre collectif, issu de la brisure spontan\'e de sym\'etrie, et ce
d\'esordre local produit par la temp\'erature. L'\'equilibre des deux \'etant
directement r\'egi par la valeur de cette m\^eme temp\'erature.

Pour mieux comprende ce ph\'enom\`ene, revenons \`a notre monde de chemises
\`a deux couleurs, o\`u la sym\'etrie a \'et\'e bris\'ee, disons vers le rouge. 
Automatiquement, plus personne alors,
n'ach\`etera de chemise verte, et tr\`es vite, le march\'e va se saturer 
avec des ventes de chemise qui vont stagner. 
Pour les faire repartir, avec des ventes de
chemises vertes, les vendeurs vont penser \`a 
baisser le prix du vert par rapport au rouge.
Et naturellement, par souci d'\'economie, cette baisse de prix 
va tenter un certain
nombre de personnes.
Mais alors, en portant leur chemise verte,
ils devront assumer une certaine tension avec 
leurs voisins
en rouge, qui 
critiqueront leur ``diff\'erence".
On peut donc imaginer
qu'une personne donn\'ee, portera sa chemise verte, seulement de temps en temps, 
pour ne pas toujours \^etre hors norme, en opposition au groupe, tout en faisant 
des \'economies.  Et le march\'e des chemises se portera mieux. 

Par contre, si le prix du vert baisse encore, la r\'ecompense de la 
marginalisation \'etant plus grande, les marginaux augmenteront automatiquement. 
Ce qui m\'ecaniquement, conduira plusieurs chemises vertes \`a se retrouver
par hasard, c™te \`a c™te, 
diminuant du m\^eme
coup, la tension sociale locale contre le vert.
Les marginaux seront donc, de 
moins en moins marginaux par leur augmentation.  \` A un certain niveau de prix 
de la chemise verte, il y aura suffisamment de gens en vert,
pour que la tension sociale disparaisse, et seul persiste l'int\'er\^et financier.
Alors tr\`es vite, tout le monde portera une chemise verte. 

Les vendeurs de chemises auront donc r\'eussi \`a changer la couleur
du groupe, mais avec de nouveau une saturation du march\'e, et en plus, avec
un profit moindre, puisque le vert est moins cher que le rouge.
Ce ph\'enom\`ene de basculement s'appelle une transition de phase 
du premier ordre. Et la diff\'erence de prix entre les couleurs 
est ce que l'on appelle 
en physique,
un champ ext\'erieur qui ``brise la sym\'etrie". Pour un syst\`eme de spins, 
ce sera un champ magn\'etique uniforme appliqu\'e dans une direction pr\'ecise.
En fait un tout petit champ suffit,
\`a d\'eterminer la couleur de tout le monde, mais cela prendra
plus de temps. Les vendeurs n'ont donc pas besoin, en fait,
de beaucoup baisser le prix du vert pour que la  nouvelle mode totalitaire 
ne devienne le vert.

Une telle situation n'\'etait pas tr\`es avantageuse pour les vendeurs de chemises.
Ils vont donc adopter une autre strat\'egie, pour \'eviter le basculement
pr\'ec\'edent. Maintenant, \`a partir de l'\'etat satur\'e, disons en rouge, 
ils vont faire des soldes, mais cette fois, simultan\'ement
sur le vert et le rouge. Cela pour ne pas briser, par les soldes, 
la sym\'etrie entre 
les deux couleurs. Ainsi la brisure spontan\'ee de la sym\'etrie initiale vers
le rouge sera maintenue malgr\'e les soldes.

Pour toucher plus de monde, ils vont \'egalement changer constamment 
les points de soldes, qui seront donc
volants. Ce sera toujours sur les deux couleurs, 
mais \`a des endroits diff\'erents, et pour un temps donn\'e.
Ainsi, au hasard de 
ces soldes mobiles, dans l'espace et dans le temps, les gens ach\`eterons,
des chemises des deux couleurs, par \'econnomie et par conformisme. 
Cependant ils porteront plus souvent le rouge que le vert, puisque 
le rouge \'etait et donc reste majoritaire. Cette fois-ci on a pas de basculement
d'une couleur dans une autre, mais bien une att\'enuation de la couleur dominante.
Il y a des fluctuations de couleurs.

Prenons un autre exemple, celui de la conduite automobile. \` A l'\'epoque
des premi\`eres voitures, il n'y avait pas de code de la route, et la circulation
se faisait selon l'envie et le bon vouloir de chacun des tr\`es rares propri\'etaires
de voitures. On roulait donc, \`a gauche, \`a droite ou au milieu. Lorsque deux 
voitures se retrouvaient face \`a face, les deux conducteurs se mettaient d'accord 
sur la fa\c con de se contouner mutuellement. 

Mais lorsque le nombre de v\'ehicules a augment\'e au point d'avoir un grand nombre
de rencontres nez \`a nez, le traffic s'en est trouv\'e interrompu, et les chauffeurs 
ont du se mettre d'accord sur un choix de circulation, soit \`a droite, soit 
\`a gauche. Le choix \'etait arbitraire quant \`a son efficacit\'e.
La France a choisit la droite, l'Angleterre la gauche, prouvant bien que l'un et
l'autre choix \'etaient possibles. Ceci est une donc une ``brisure de la sym\'etrie" 
qui existait initialement, pour quelques voitures isol\'ees, 
mais qui a \'et\'e supprim\'e au profit
d'une efficacit\'e collective. On peut sch\'ematiser la situation par 
les voies \`a ligne jaune continue. 

Ensuite les voitures ont commenc\'e a avoir des vitesses tr\`es diff\'erentes, et 
l'impossibilit\'e de doubler est devenue un frein \`a l'efficacit\'e pr\'ec\'edente.
On est alors pass\'e \`a la ligne jaune pointill\'ee, qu'on peut franchir 
mais pour un temps court. On a ainsi r\'etabli la sym\'etrie qui existait
initialement, mais de fa\c con sporadique et transitoire, pour de nouveau
augmenter l'efficacit\'e du syst\`eme. on a donc l'existence d'un certain d\'esordre
dans la conduite.

La question se pose alors de savoir, jusqu'o\`u peut aller l'introduction 
de ce d\'esordre local, dans l'ordre initial. 
En termes physiques, que se passe-t-il 
lorsqu'on augmente de plus en plus la temp\'erature ? 

Dans le cas des chemises, l'analogue de la temp\'erature est l'amplitude des
soldes simultan\'ees sur le rouge et le vert, par rapport
au prix de r\'ef\'erence.
Pour de petites baisses, les gens porteront plus souvent, la couleur 
dominante. Mais plus la baisse sera importante, 
plus ils porteront les deux couleurs indifferemment, et aussi souvent 
l'une que l'autre. Les fluctuations de couleurs auront alors
\'enorm\'ement augment\'ees. Dans ce cas,
le syst\`eme subi encore, une transition de phase, mais cette
fois, de nature diff\'erente de la pr\'ecedente, car ici, la sym\'etrie initialement
pr\'esente entre les couleurs, a \'et\'e r\'etablie collectivement.
Ce type de transition est qualifi\'ee du deuxi\`eme ordre, par rapport 
\`a la pr\'ecedente, dite du premier ordre. Dans le premier cas, 
on a eu, \`a un certain moment, un basculement brutal de l'\'etat collectif 
du syst\`eme, alors que
pour une transition
du deuxi\`eme ordre, le changement se fait de fa\c con continue, 
mais avec beaucoup de fluctuations.

Apr\`es la transition, on a une r\'epartition \'equilibr\'ee des couleurs. 
C'est ce qu'on appelle en
physique une phase d\'esordonn\'ee, c'est \`a dire une phase, o\`u la brisure
de sym\'etrie a disparue. Mais attention dans une telle phase, les corr\'elations
de choix de couleur, entre les individus
n'ont pas disparues pour autant. Il reste autant de chemises 
vertes que de rouges, avec toujours une tendance local \`a l'uniformisation,
mais qui maintenant ne se propage plus \`a toute la population. On a alors
un d\'esordre maximum.
Par contre, dans le cas des voitures, trop augmenter le d\'esordre produirait 
de plus en plus d'accidents, et bloquerait toute la circulation.

Pour diff\'erencier une phase ordonn\'ee avec du d\'esordre,
(mais o\`u la brisure de sym\'etrie persiste), 
d'une phase d\'esordonn\'ee (sans sym\'etrie bris\'ee),
il suffit de renverser les ``couleurs" respectives de chacun. 
Si la couleur d'ensemble du groupe a chang\'ee (pass\'ee d'un exc\'es de vert 
\`a un exc\'es de rouge,
ou vice versa), c'est que la sym\'etrie est bris\'ee. Par contre si elle
reste globalement
la m\^eme (mi-vert, mi-rouge) alors la sym\'erie n'est pas bris\'ee.

On peut d'ailleurs mesurer le degr\'e de la brisure de sym\'etrie
par un param\`etre,
le param\`etre d'ordre.
Il est est \'egal \`a un, l'ordre est total, totalitaire,
lorsque on est \`a temp\'erature nulle. Et il va diminuer vers z\'ero
en fonction de l'augmentation de temp\'erature.
 Il va d'abord d\'ecro"tre lentement, puis
de plus en plus vite, pour finalement s'annuler,
\`a une certaine valeur $T_c$ de la temp\'erature, appel\'ee
la temp\'erature critique  du syst\`eme. Elle varie d'un
syst\`eme \`a un autre. Au del\`a de la temp\'erature critique $T_c$, 
le param\`etre d'ordre reste nul, ce quelle que soit la valeur de la temp\'erature.

Dans le cas des chemises, le param\`etre d'ordre est naturellement 
\`egal au nombre de chemises rouges moins le nombre de chemises vertes,
divis\'e par le nombre total d'individus.
\` A prix \'egal de couleur, ce nombre est donc \'egal \`a $+1$. Une baisse 
uniquement du vert le ram\`ene \`a $-1$. Par contre des soldes mobiles et volantes 
le font d\'ecro"te de $+1$ vers z\'ero, qu'il atteint dans la phase d\'esordonn\'ee,
pour une certaine baisse de prix.

\` A des temp\'eratures $T$ inf\'erieures \`a $T_c$, 
pour des transitions du deuxi\`eme ordre, le param\`etre
d'ordre se comporte proportionnelement \`a une puissance de la distance 
en tem\'erature  \`a $T_c$. Il est proportionnel \`a
$(T_c-T)^\beta $, o\`u $\beta$ est appel\'e
un exposant critique. 
Il a \'et\'e remarquable de constater, et de prouver,
que ce comportement en loi de puissance
est universel. En effet, la valeur de $\beta$ est identique pour
un grand nombre de syst\`emes physiques de nature tr\`es diff\'erente, comme par 
exemple un liquide \`a son point d'\'ebullition, ou un aimant qui chauffe au point
de perdre son aimantation. Par contre la valeur de $T_c$, elle, varie
 d'un syst\`eme \`a l'autre.
Ce caract\`ere d'universalit\'e
montre que ce qui est en jeu, r\'eside dans l'aspect ``comportement collectif" 
du syst\`eme,
et non pas dans ses ``propri\'et\'es intrins\`eques" comme par exemple
la nature de ces interactions. Certains physiciens tentent
d'ailleurs d'\'etendre cette universalit\'e des transitions de phase 
\`a certaines classes de ph\'enom\`enes sociaux et \'economiques.

Au point tr\`es pr\'ecis o\`u le param\`etre d'ordre s'annule, il se passe des choses 
surprenantes. On a y trouv\'e en effet
que chaque \'el\'ement du syst\`eme est corr\'el\'e, 
c'est \`a dire en communication, 
avec tous les autres \'elements du m\^eme syst\`eme, et cela
m\^eme si ce dernier est 
de taille infinie. Un individu qui change de couleur va donc influencer tous
les autres, et vice versa.
Cela \`a tel point que pour un individu donn\'e, la situation
pourrait sembler comme ``mystique" (c'est une image).

On assiste alors \`a des fluctuations g\'eantes de couleur, qui sont 
simultan\'ees avec
des fluctuations minuscules. Cette coexistence d'une multitude d'\'echelles
de longueur produit, ce que l'on appelle 
une ``invariance d'\'echelle".
Quel que soit le niveau auquel on regarde ce qui se passe, c'est toujours 
exactement identique. C'est comme si on regardait un paysage \`a l'oeil nu, 
au microscope, ou avec un zoom g\'eant, et que l'image ne changeait pas, ce
qui est tout m\^eme extraordinaire. Ce sont par exemple, 
les fluctuations de densit\'e, \`a toutes les \'echelles
d'un liquide \`a son point critique, qui donnnent
le ph\'enom\`ene experimental de 
l'opalescence critique, o\`u de la lumi\`ere envoy\'ee sur le liquide
est si fortement diffus\'ee, qu'il semble tout entier s'illuminer.

De plus, lorsque la temp\'erature s'approche du point critique, 
qu'elle lui soit sup\'erieure, ou inf\'erieure, 
le syst\`eme dans son ensemble r\'eagit massivement \`a toute
perturbation ext\'erieure qui voudrait briser sa sym\'etrie. 
Par exemple, un tout petit champ magn\'etique
va aligner dans sa direction tous les spins d'un morceau de fer.
Autrement dit, le nombre d'individus suceptible de r\'eagir \`a une petite
perturbation ext\'erieure qui brise la sym\'etrie,
est infinie. Il y a donc ce que l'on appelle
une divergence de la fonction r\'eponse du syst\`eme,
en l'ocurrence, sa suceptibilit\'e devient infinie au point critique. 
Elle diminue de part et d'autre de $T_c$.

L\`a encore on retrouve
une propri\'et\'e d'universalit\'e similaire \`a celle du param\`etre d'ordre.
sur la fa\c con dont se fait cette divergence.
Au voisinage de $T_c$, et de part  et d'autre, la suceptibilit\'e 
diverge comme $(T_c-T)^{-\gamma }$. La valeur de l'exposant $\gamma $, 
diff\'erente de celle de $\beta$,
est aussi comme elle, universelle, c'est \`a dire qu'elle
est la m\^eme pour toute une classe de syst\`emes physiques de natures diff\'erentes.

La vente des chemises s'int\'egrant ou non dans la classe d'universalit\'e
des syst\'emes magn\'etiques est une question encore ouverte, qui 
d'ailleurs n'a pas \'et\'e pos\'ee.
Et donc, l'objectif de cet article n'\'etait pas d'y r\'epondre, mais plut™t
de faire saisir
quelques m\'ecanismes essentiels des transitions de phase, \`a partir d'une
petite m\'etaphore sociale.

%%%%%%%%%%%%%%%%%%%%%%%
\end{document}